\documentclass{article}
\usepackage{arxiv}

\usepackage[utf8]{inputenc} % allow utf-8 input
\usepackage[T1]{fontenc}    % use 8-bit T1 fonts
\usepackage{hyperref}       % hyperlinks
\usepackage{url}            % simple URL typesetting
\usepackage{booktabs}       % professional-quality tables
\usepackage{amsfonts}       % blackboard math symbols
\usepackage{nicefrac}       % compact symbols for 1/2, etc.
\usepackage{microtype}      % microtypography
\usepackage{lipsum}		% Can be removed after putting your text content
\usepackage{graphicx}
\usepackage{natbib}
\usepackage{doi}
\usepackage{siunitx}

\title{BraggNN: Fast X-ray Bragg Peak Analysis Using Deep Learning}

\newcommand{\proj}{\texttt{BraggNN}}
\newcommand{\nfhedm}{\texttt{nf-HEDM}}

\author{Zhengchun Liu \\
	Data Science and Learning division\\
	Argonne National Laboratory\\
	Lemont, IL 60564 \\
	\texttt{zhengchun.liu@anl.gov} \\
	
	\And
    Hemant Sharma\\
    X-ray Science Division\\
    Argonne National Laboratory\\
    Lemont, IL 60439 \\
    \texttt{hsharma@anl.gov} \\
    
    \And
    Jun-Sang Park\\
    X-ray Science Division\\
    Argonne National Laboratory\\
    Lemont, IL 60439 \\
    
    \And
    Peter Kenesei \\
    X-ray Science Division\\
    Argonne National Laboratory\\
    Lemont, IL 60439 \\
    \And
    Antonino Miceli \\
    X-ray Science Division\\
    Argonne National Laboratory\\
    Lemont, IL 60439 \\
    \And
    Jonathan Almer \\
    X-ray Science Division\\
    Argonne National Laboratory\\
    Lemont, IL 60439 \\
    
    \And
    Rajkumar Kettimuthu \\
	Data Science and Learning division\\
	Argonne National Laboratory\\
	Lemont, IL 60564 \\
	
	\And
    Ian Foster \\
	Data Science and Learning division\\
	Argonne National Laboratory\\
	Lemont, IL 60564 \\
    
}

\date{}

\renewcommand{\shorttitle}{BraggNN: Fast X-ray Bragg Peak Analysis Using Deep Learning}

\begin{document}
\maketitle

\begin{abstract}
X-ray diffraction based microscopy techniques such as High Energy Diffraction Microscopy rely on knowledge of the position of diffraction peaks with high precision. These positions are typically computed by fitting the observed intensities in area detector data to a theoretical peak shape such as pseudo-Voigt. As experiments become more complex and detector technologies evolve, the computational cost of such peak detection and shape fitting becomes the biggest hurdle to the rapid analysis required for real-time feedback during in-situ experiments.
To this end, we propose \proj{}, a deep learning-based method that can determine peak positions much more rapidly than conventional pseudo-Voigt peak fitting.
When applied to a test dataset, \proj{} gives errors of less than 0.29 and 0.57 pixels, relative to the conventional method, for 75\% and 95\% of the peaks, respectively.
When applied to a real experimental dataset, a 3D reconstruction that used peak positions computed by \proj{} yields 15\% better results on average as compared to a reconstruction obtained using peak positions determined using conventional 2D pseudo-Voigt fitting.
Recent advances in deep learning method implementations and special-purpose model inference accelerators allow \proj{} to deliver enormous performance improvements relative to the conventional method, running, for example, more than 200 times faster than a conventional method on a consumer-class GPU card with out-of-the-box software.
\end{abstract}

% keywords can be removed

\section{Introduction}
Advanced materials affect every aspect of our daily lives, including the generation, transmission and use of energy.
Accelerating the pace of materials design promises to enhance economic activity and the transition to a cleaner energy future. 
However, current material design approaches rely heavily on intuition based on past experiences and empirical relationships. 
In order to qualify new materials for critical applications, several high-energy X-ray characterization methods have been developed over the past decade. 
One of the foremost is high-energy diffraction microscopy (HEDM) \citep{park2017far}, which provides non-destructive 3D information on structure and its evolution within  polycrystalline materials. 
HEDM techniques have enabled breakthroughs in understanding of various processes, through carefully designed experiments that are tractable for analysis by researchers \citep{NARAGANI201771,bernier2020,WANG2020152534}.
These methods use diffraction and tomographic imaging of up to cm-sized objects with resolutions down to the micrometer level.

\begin{figure}
\centering
\includegraphics[width=.98\linewidth,bb=0 0 420 360]{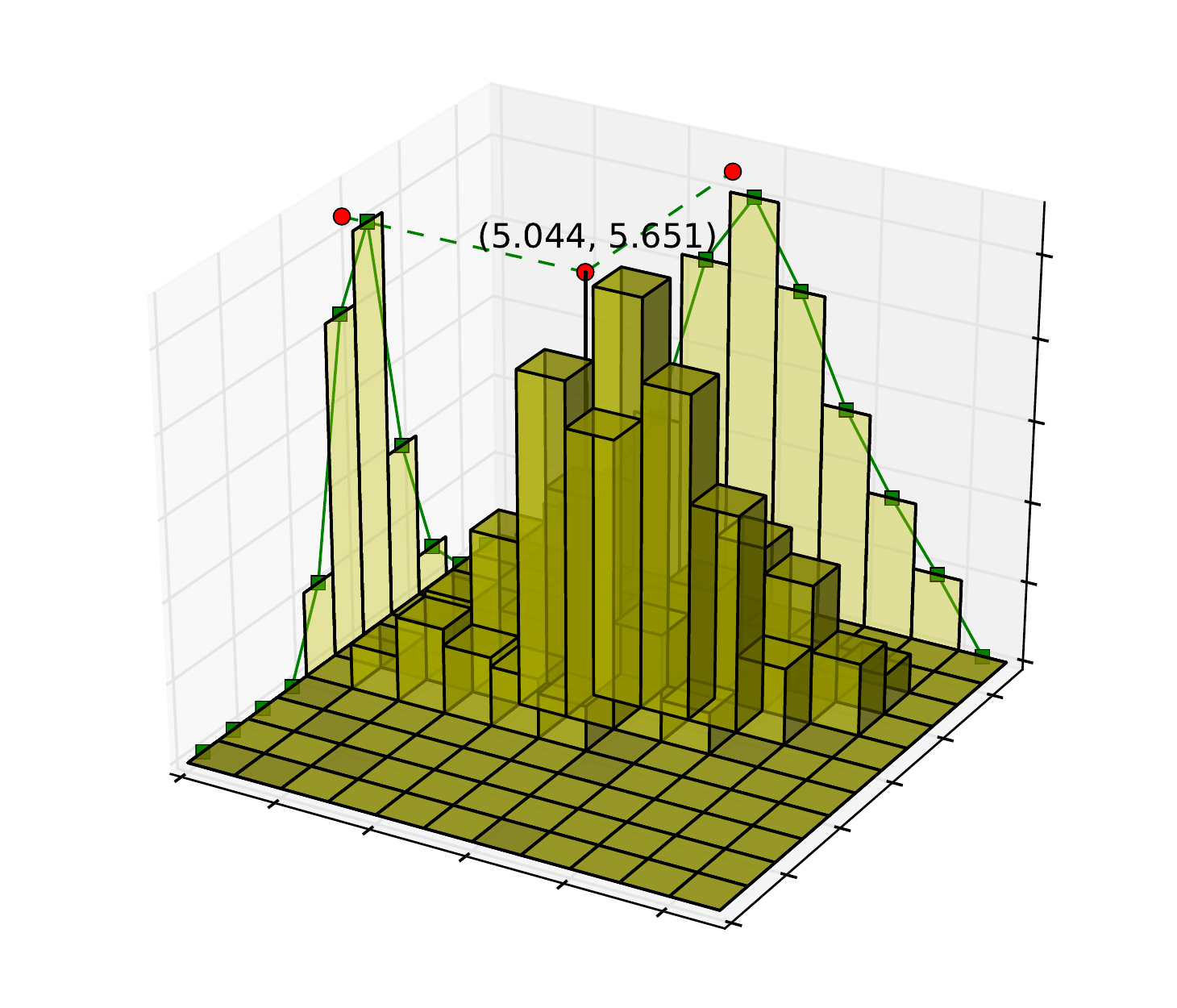} 
\caption{A diffraction peak in X-ray diffraction ($11\times11$ patch). The height denotes photon counts, and the red dots show the peak position computed by fitting a pseudo-Voigt profile.}
\label{fig:3dpeak}
\end{figure}

A conventional HEDM experiment involves four steps: (1) data acquisition, (2) transfer of full scan from detector to central storage, (3) offline Bragg peak analysis to determine precise peak characteristics, and (4) reconstruction of grain information from the peak positions generated in the third step \citep{Sharma2012ASetup,Sharma2012AGrains}.
A single typical HEDM scan involves illuminating the polycrystalline aggregate of interest and acquiring diffraction images (1440--3600 frames in total) while rotating the specimen at a constant speed.
Data acquisition is increasingly fast: a single typical HEDM scan consisting of 1440--3600 frames takes about 6--15 minutes to acquire today at the Advanced Photon Source (APS) and projected to be 50--100 seconds after the planned upgraded APS (APS-U)~\citep{APSU} with faster detectors. 
Rotation of the specimen enables each grain to satisfy the Bragg-diffraction condition multiple times, resulting in multiple diffraction peaks from the grain.
Reconstruction of Far-Field (FF) HEDM data depends on determination of the peak positions with sub-pixel accuracy, which can deviate significantly from the maxima as shown in Figure \ref{fig:3dpeak}.

Peak positions are typically computed by (optionally) transforming the peaks to polar coordinates and then fitting the peaks to a pre-selected peak shape such as Gaussian, Lorentzian, Voigt, or Pseudo-Voigt \citep{Sharma2012ASetup}.
The Voigt profile, a probability distribution given by a convolution of a Cauchy-Lorentz distribution and a Gaussian distribution, is often used in analyzing data from spectroscopy or diffraction. 
However, it is computationally expensive to fit a Voigt profile in 2D (or 3D) space for each Bragg peak, so the peak shape is approximated to a pseudo-Voigt profile.
Depending on sample properties and the extent of the mechanical, thermal, electromagnetic, or chemical stimuli applied to the sample, processing time can range from 10 minutes to a few weeks for one typical HEDM dataset, even when using an HPC cluster with thousands of CPU cores. 
These long data analysis times are more than just an inconvenience: they prevent experimental modalities that depend on measurement-based feedback for experiment steering.

Although we describe \proj{} framework as applied to FF-HEDM, it is also useful for other diffraction techniques dealing with single or polycrystal diffraction. The data and source code that support the findings of this study are openly available at \url{https://github.com/lzhengchun/BraggNN}.

\section{\proj{} and its Training}
A significant fraction of HEDM data analysis time is spent on determining peak positions, motivating us to seek methods for accelerating this operation.
Artificial neural networks are known for their universal approximation capability that allows them to represent complex and abstract functions and relationships~\citep{hornik1991approximation}.
Thus, a promising solution to the Bragg peak localization challenge is to train a deep learning model to directly approximate the position of Bragg peaks. 
Advances in both machine learning~(ML) methods and AI inference accelerators allow such a model to run far faster than conventional methods, making it feasible to extract peak information from streaming data in real-time, enabling rapid feedback and reducing downstream transfer, storage, and computation costs.

\subsection{Model Design}\label{sec2:model-design}
Deep learning~(DL) is part of a broader family of machine learning methods based on artificial neural networks to progressively extract higher level features from the pixel-level input through a hierarchical multi-layer framework.
The convolutional neural network~(CNN), a widely used building block of DL models for visual image analysis, is parameter efficient due to the translation-invariant property of its representations, which is the key to the success of training deep models without severe over-fitting. 
Although a strong theory is currently missing, much empirical evidence supports the notion that both the translation-invariant property and convolutional weight sharing (whereby the same weights are shared across an entire image) are important for good predictive performance~\citep{cheng2017survey}. 

In this work, we express this task as a regression problem using supervised machine learning and present \proj{}, a deep neural network-based model, for precisely localizing Bragg peaks far more rapidly than that can be achieved by applying conventional fitting methods to peak shape profiles.
Note that we are not concerned here with the problem of locating within an image a patch that contains a peak (the \emph{object localization problem}) because Bragg peaks are easily separated from background by using a heuristic thresholding value, and from neighbor peaks by using a connected-component labeling algorithm (overlapped peaks and fitting in 3D will be studied in future work)~\citep{fiorio1996two,wu2005optimizing}.
Our problem rather is to determine, with sub-pixel precision, the center-of-mass of a diffraction peak in a supplied patch: the \emph{peak localization problem}.

\begin{figure}
\centering
\includegraphics[width=\linewidth]{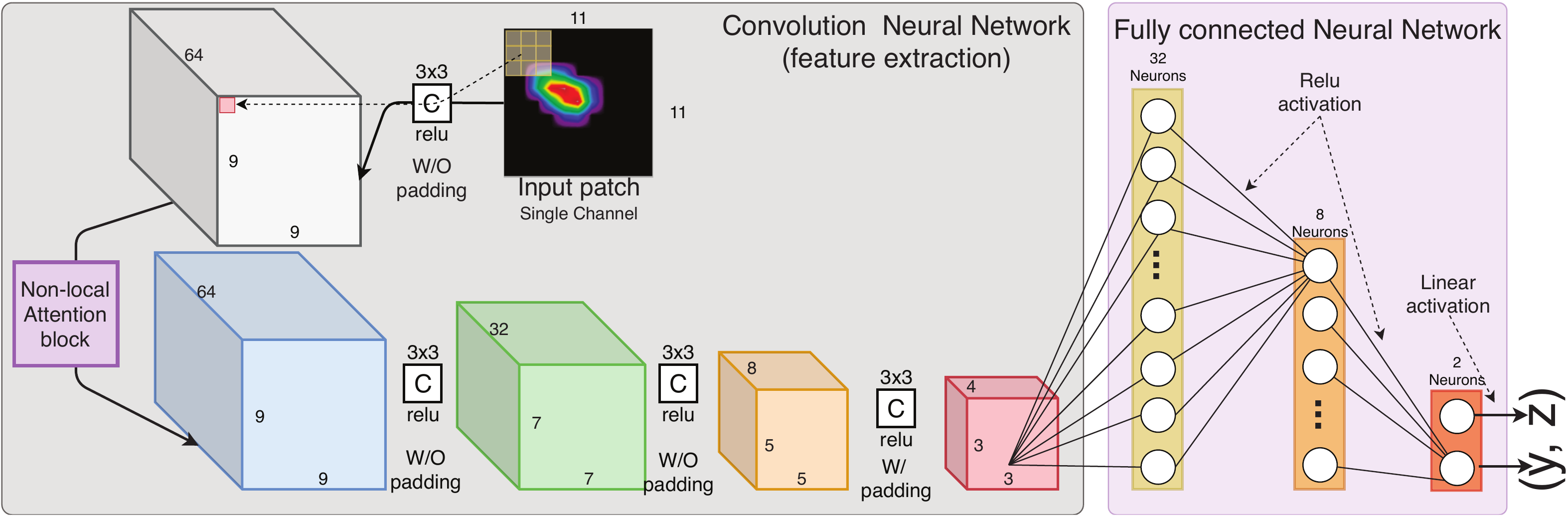}
\caption{Application of the \proj{} deep neural network to an input patch yields a peak center position (y, z). All convolutions are 2D of size $3\times3$, with rectifier as activation function. Each fully connected layer, except for the output layer, also has a rectifier activation function.}
\label{fig:hedm-nnv2}
\end{figure}

The \proj{} network architecture, shown in Fig.~\ref{fig:hedm-nnv2},  comprises  series of CNN layers (four in the figure) acting as feature extractors are followed by a series of fully connected layers (three in the figure) that generate a regression prediction. 
Each CNN kernel/filter is an artificial neuron that, in contrast to traditional algorithms in which kernels are hand-engineered, learns to extract a type of feature (e.g., various oriented edges, or blobs of color) from its input. 
Each 2D CNN neuron has $3 \times 3 \times c$ learn-able weights plus one learn-able bias to convolve a feature map (a 3D volume shaped as height $\times$ width $\times$ depth/channel) with $c$ channels (e.g., the input patch has one channel as shown in the figure).

Here we use the first layer, which takes a Bragg peak in a patch with $11 \times 11 \times 1$ ($c=1$) pixels as input and outputs 64 feature maps (each has $9\times 9$ pixels), as an example to explain the convolution operation.
At every convolution position, for example the one shown as a dotted line in the figure, the dot product between the weights and the input entry~($3 \times 3 \times c$ centered at the convolution position) is computed and added to the learn-able bias.
%Convolutional networks were inspired by biological processes in that the connectivity pattern between neurons resembles the organization of the animal visual cortex. Individual cortical neurons respond to stimuli only in a restricted region of the visual field known as the receptive field. The receptive fields of different neurons partially overlap such that they cover the entire visual field.
This convolution result, called the activation, is then passed through a rectified linear unit~(ReLU, $f_{relu(x)}=max\left(x, 0\right)$) activation function to yield a feature.
Each kernel is convolved~(vertically and horizontally) across the input image, producing a $9 \times 9$ feature map. 
Thus, although the operation is colloquially referred to as a convolution, mathematically, it is a sliding dot product or cross-correlation.
Each layer has multiple independent neurons that result in multiple feature maps. 
All feature maps are stacked along the depth dimension and then passed to the next layer as input.
For example, as the first layer has 64 neurons, it turns a $11 \times 11 \times 1$ input patch to a $9 \times 9 \times 64$ volume.  
Multiple convolutions layers are chained to encode the input image into a representation in latent space. 
%We note that, by design we padded zeros around the dimension of height and width to the input feature volume before feeding into the last CNN layer so as to get an output volume with the same width and height as its input.

The fully connected (FC) neural network layer takes the encoded representation produced by the CNN layer as input, estimates the center of the input Bragg peak, and produces the (x, y) coordinates as output.
In a similar manner to the CNN layer, each FC layer has multiple artificial neurons, each of which has the same number of learn-able weights as its input plus one learn-able bias. 
The 3D feature map~(e.g., $5 \times 5 \times 4$) produced by the last CNN layer is reshaped into a 1D vector before feeding it into the first FC layer.
The dot product between the neuron's weights and input is computed and added to the bias. 
Thus, $n$ neurons in a given FC layer generate an output vector of  dimension $n$, which are passed into the ReLU activation function and then serve as the input of the next layer.
As one can see, unlike the CNN neurons that receive input from only a restricted subarea of the previous layer for each convolution point, each neuron in a FC layer is connected to all neurons in the previous layer.
The output layer in our design has no activation function~(or, equivalently, it applies a linear activation). 

The whole process that turns an input Bragg peak patch into two floating point numbers~(the coordinates of the peak center) is called a feed forward pass. 
The $\ell_2$-norm is computed between the model output and ground truth~(estimated by using pseudo-Voigt fitting) as the model loss.
Training then proceeds as follows.
We compute the gradient of each neuron's weights with respect to the loss function by using the chain rule~(implemented via automatic differentiation in deep learning frameworks such as PyTorch~\citep{paszke2019pytorch}). 
This process of computing the gradient of learn-able weights is called back propagation. 
The neuron's weights are then updated by using the gradient descent optimization algorithm~\citep{kingma2014adam}.
Training iterates the feed-forward and back-propagation process on different (Bragg peak patch, ground truth center) pairs many times, until the model no longer makes noticeable progress in minimizing the $\ell_2$-norm.

We train the \proj{} model with a collection of input-output pairs, each with a peak patch as input and the peak center position obtained from pseudo-Voigt fitting as output.
Once the \proj{} model is trained, we can then apply it to patches obtained from new diffraction data as follows:
(1) We use the connected-component labeling algorithm~\citep{fiorio1996two,wu2005optimizing} to detect connected regions (i.e., peaks) in binary digital images. If the region has multiple maxima, indicating the presence of overlapping peaks, the region is discarded. Overlapping peaks will be investigated in a later study.
(2) For each region detected in the previous step, we determine the coordinate (row and column index of the image matrix) of its peak (maxima), and crop a patch with a pre-defined size (an odd number, must be the same as that used for training \proj{}) with the peak coordinate as the geometric center.
Application of the trained \proj{} model to such a patch then yields an estimate of the peak position. 
Given this center of mass, we then map the position of the peak in the patch back to the diffraction frame based on the location of the patch in the diffraction frame. 
Each diffraction frame is processed independently, focusing only on 2D shapes of the peaks. In case of heavily deformed materials, where orientation changes within grains cause diffraction signal to be present in multiple frames, 3D peak processing would be required and will be investigated in a future study.

\subsection{Data Augmentation}\label{sec2:data-aug}
The performance of a deep neural network depends heavily on the quantity, quality, and diversity of the data used to train the model.  
If data are not sufficiently diverse, it is easy to experience overfitting, whereby a network learns a function with very high variance that models the training data perfectly but performs badly on other data. 
Many application domains, including ours, lack access to large (in terms of both quantity and diversity) and high-quality (accurately annotated) labelled data.
Data augmentation is a strategy that enables practitioners to significantly increase the diversity of data available to train their DL models, without actually collecting new data. 
Data augmentation techniques such as cropping, padding, and horizontal flipping are commonly used to train large neural networks for image classification~\citep{AutoAugment,shorten2019survey} such as CIFAR-10~\citep{krizhevsky2009learning} and ImageNet~\citep{deng2009imagenet}.

While some existing data augmentation techniques may be useful in the Bragg peak context to avoid over-fitting, none are useful for training a more generic model (e.g., one that generalizes to data outside the training set, or that handles unseen peaks cropped from noisy diffraction frames) because the augmented samples used in the above mentioned techniques will not be found in practice.  

\begin{figure}
\centering
\includegraphics[width=0.98\linewidth]{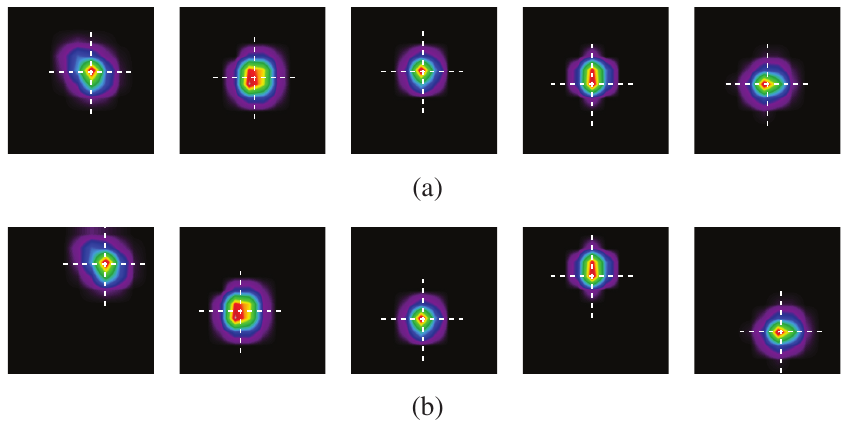}
\caption{Demonstration of a mini-batch of patches~(11$\times11$ pixels) with data augmentation for model training. (a) Peak maxima is in the center of the patch. (b) Peak maxima is intentionally offset by up to $\pm$2 pixels in horizontal and vertical direction for the same peaks.}
\label{fig:data-augu-demo}
\end{figure}

Thus, we introduce a novel physics-inspired data augmentation method that can both avoid overfitting and help to train a more generic model able to deal with imperfect peak cropping from noisy diffraction frames.
Specifically, when cropping patches for model training we deviate the peak center from the geometric center randomly by up to $\pm$2 pixels in horizontal and vertical directions.
%$m \in \mathbb{Z}$ and $n \in \mathbb{Z}$ pixels in the horizontal (positive to right, negative to left) and vertical (positive for down, negative for up) directions independently. 
Fig.~\ref{fig:data-augu-demo} demonstrates a batch of 10 patches with (Fig.~\ref{fig:data-augu-demo}(a)) and without (Fig.~\ref{fig:data-augu-demo}(b)) data augmentation.

This data augmentation approach helps to train a more general model because, like regularization, \citep{zhang2016understanding} it adds prior knowledge~(the center-of-mass is not always near the geometric center) to a model training and increases training data. 
It also helps to make the testing dataset statistically more similar to the peak patches that will be encountered during inference in production. 
The ablation evaluation in Appendix \ref{appsec2:augm-ablation} shows the effectiveness of this novel data augmentation method.

\subsection{Model Training}\label{sec2:model-training}
An important tunable parameter when training a model is the input patch size, as shown in Fig.~\ref{fig:hedm-nnv2}.
The appropriate patch size depends on the pixel size of the detector and size of the diffraction peaks.
Best practice is to choose a patch size that can fully cover all valid peaks and still leave 2--4 pixels from peak edge to patch edge for data augmentation.
Since the input patch size will determine the size of the neurons in the first fully connected layer, a model trained with one patch size cannot work with another patch size in practice. 
Another tunable parameter is for data augmentation, i.e., the interval of $m$ and $n$. 
We typically choose the same interval size for $m$ and $n$.
During training, we independently sample (with replacement) a number from the interval for $m$ and $n$ separately in order to prepare each sample of each mini-batch online. 

We implemented our model by using the PyTorch~\citep{paszke2019pytorch} machine learning framework.
We train the model for a maximum of \num{80000} iterations with a mini-batch size of 512, with validation-based early stopping applied to avoid using an over-fitted model for testing and production use. 
Training takes about two hours using one NVIDIA V100 GPU, and only a few minutes using the Cerebras CS-1\footnote{https://www.cerebras.net/} artificial intelligence computer.

In the experimental studies reported in this paper we train and evaluate \proj{} on a diffraction scan dataset collected using an undeformed bi-crystal Gold sample \citep{ShadeGold} with 1440 frames (0.25$^\circ$ steps over 360$^\circ$) totaling \num{69347} valid peaks.
We use 80\% of these peaks (\num{55478}) as our training set, \num{6000} peaks ($\sim$9\%) as our validation set for early stopping \citep{goodfellow2016deep} and the remaining \num{7869} peaks ($\sim$11\%) as a test dataset for model evaluation. %, as will be discussed in \S\ref{sec:mdl-eva}.

\section{Results Analysis and Discussion}
Once the model is trained, we evaluate its performance from two perspectives: 1) we measure distance~(i.e., error) between each \proj{}-estimated center and the corresponding center obtained via conventional pseudo-Voigt profile (conventional method); 2) we apply \proj{} to an experiment of a different sample, reconstruct grains using peak information by \proj{}, and compare reconstructed grain size and position with those reconstructed using conventional method~\citep{Sharma2012ASetup}.

\subsection{Model Performance}\label{sec2:eva-peak}

We start with quantitatively evaluating \proj{} by looking at the accuracy of Bragg positions estimated by it.

\begin{figure}
\centering
\includegraphics[width=0.98\linewidth]{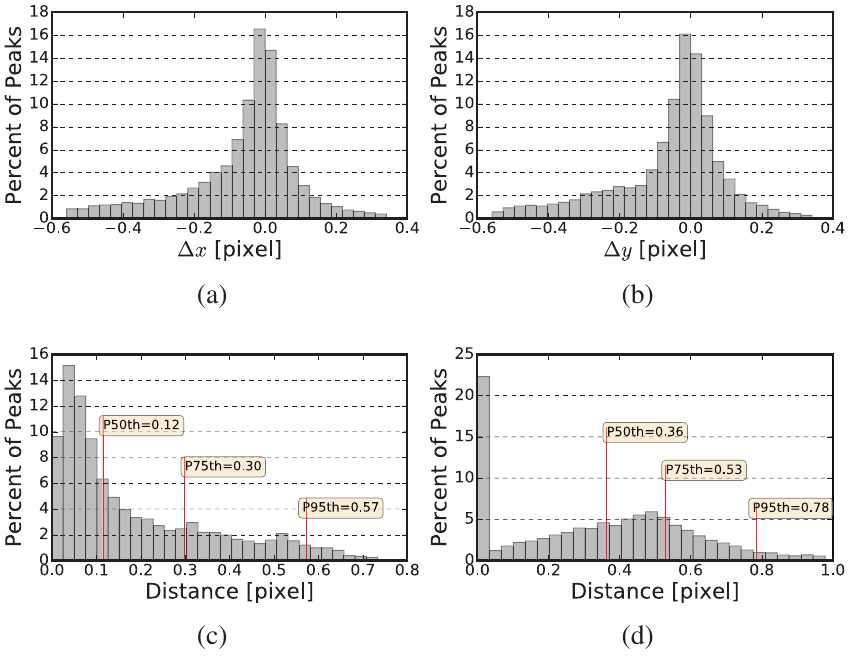}
\caption{Distribution of difference between peak positions determined using \proj{}(a-c) or Maxima(d) and the conventional pseudo-Voigt fit. (a) Difference in horizontal direction, (b) difference in vertical direction, (c) euclidean distance between peak position, and (d) euclidean distance between peak positions determined using maxima position of peaks and conventional pseudo-Voigt fit. $P_{n}$ in (c) and (d) denotes the Euclidean distance at the $n^{th}$ percentile.}
\label{fig:peak-diff}
\end{figure}

Fig.~\ref{fig:peak-diff} shows the distribution of difference between position of diffraction peaks determined using \proj{} (a-c) or peak Maxima (d) and conventional pseudo-Voigt fit.
Fig.~\ref{fig:peak-diff} (a-b) show that \proj{} tends to underestimate the positions more often than overestimating them.
As quantified using Euclidean distance in Fig.~\ref{fig:peak-diff}(c), most peaks deviate little (e.g., 75\% of peaks deviate less than 0.3 pixels) from the position identified by conventional method. 
The difference between position of the maximum intensity of Bragg peaks (with resolution of up to 0.5 pixels) and conventional method results, shown in Fig.~\ref{fig:peak-diff}(d), are much higher than differences between \proj{} and conventional method results Fig.~\ref{fig:peak-diff}(c).

\subsection{Reconstruction Error Analysis}\label{sec2:eva-recon}

Since the reconstruction of grain positions is our final goal, we also evaluate the trained \proj{} on a different dataset~\citep{turner2016combined} and compare with results from the conventional method.
The dataset consists of Far-Field (FF) and Near-Field (NF) HEDM data acquired \textit{in-situ} during deformation of a Ti-7Al sample.
FF-HEDM data was acquired using the same beam configuration at the same location in the sample as NF-HEDM data, thus enabling a one-to-one comparison of Center-Of-Mass (COM) position of grains.
FF-HEDM directly outputs COM position of grains, whereas COM positions were calculated from NF-HEDM reconstructions using the voxelized information.
A single 2D slice of the specimen was reconstructed using the MIDAS software package \citep{Sharma2020,Sharma2012AGrains,Sharma2012ASetup} in three different configurations: FF-HEDM reconstruction using \proj{}, FF-HEDM reconstruction using conventional pseudo-Voigt fitting and NF-HEDM reconstruction. 

\begin{figure}
\centering
\includegraphics[width=0.98\linewidth]{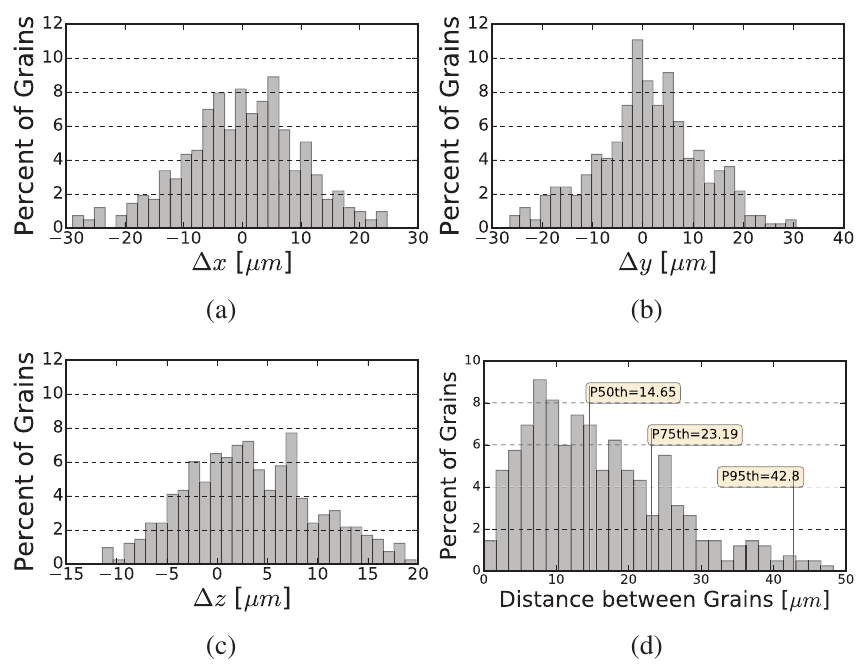}
\caption{Distribution of position differences between grains when reconstructed by using \proj{} peak position and by pseudo-Voigt fitting. (a) x-axis, (b) y-axis, (c) z-axis, (d) Euclidean distance. $P_{n}$ in (d) denotes the $n^{th}$ percentile.}
\label{fig:codl-diff}
\end{figure}

First we compare the FF-HEDM reconstructions using peak positions obtained from \proj{} to the FF-HEDM reconstructions using conventional pseudo-Voigt fitting, shown in Fig.~\ref{fig:codl-diff}. The difference in position in the $x$-axis (along the x-ray beam, Fig.~\ref{fig:codl-diff}(a)), $y$-axis (horizontal direction perpendicular to the $x$-axis, Fig.~\ref{fig:codl-diff}(b)) and $z$-axis (vertical direction coincident with the rotation axis, Fig.~\ref{fig:codl-diff}(c)) are centered around 0, implying there is no systematic bias between the two reconstructions. The euclidean distance between grains reconstructed using \proj{} and conventional pseudo-Voigt fitting (Fig.~\ref{fig:codl-diff}(d)) is less than 15 $\mu$m for 50\% of the grains.

\begin{figure}
\centering
\includegraphics[width=\linewidth]{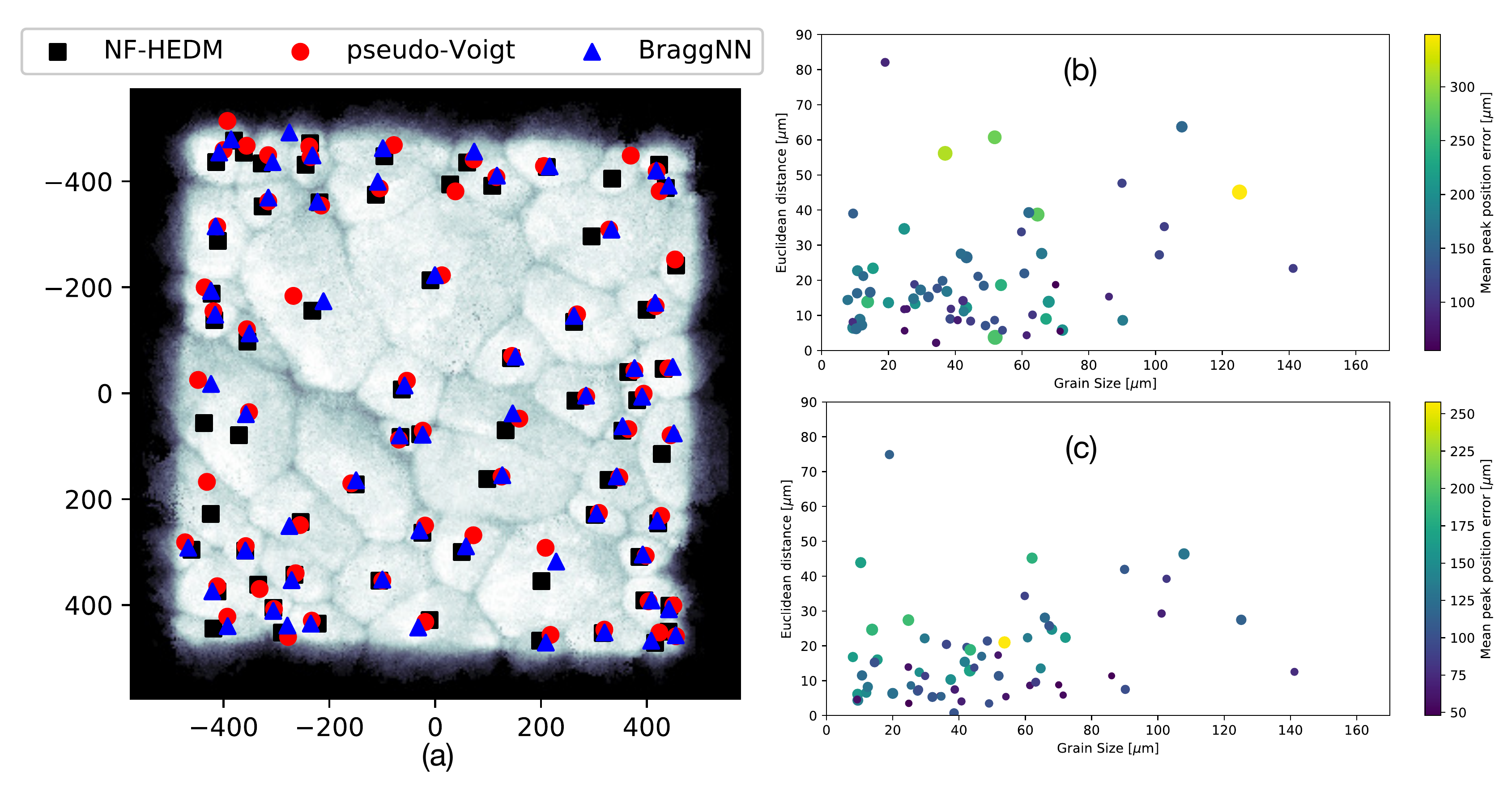}
\vspace{-5mm}
\caption{A comparison of \proj{}, pseudo-Voigt FF-HEDM and NF-HEDM. (a) Grain positions from NF-HEDM (black squares), pseudo-Voigt FF-HEDM (red circles) and \proj{} FF-HEDM (blue triangles) overlaid on NF-HEDM confidence map. (b-c) Difference in position of grains between pseudo-Voigt FF-HEDM (b), \proj{} (c) and NF-HEDM as a function of Grain Size. Color of markers in (b-c) represent the mean difference in position of expected and observed diffraction spots. Size of markers in (b-c) represent the mean Internal Angle (see text).}
\label{fig:nf-hedm-cmp}
\end{figure}

We note that, although we compare \proj{} with conventional pseudo-Voigt fitting, the ill-posed inverse problem means that conventional pseudo-Voigt is not the ground truth. 
Therefore, we use and compare the grain COMs estimated from NF-HEDM, which results in higher resolution reconstructions by providing a space-filling orientation map, to grain COMs obtained using the FF-HEDM reconstruction technique that is the focus of this paper.
A total of 68 grains were identified using NF-HEDM, out of which all the grains could be matched using conventional pseudo-Voigt fitting, but 6 grains were not detected using \proj{} because \proj{} ignored overlapping peaks.
Fig.~\ref{fig:nf-hedm-cmp}(a) shows the position of grains imaged using the three different reconstruction methods (NF-HEDM, pseudo-Voigt FF-HEDM and \proj{} FF-HEDM) overlaid on grain shapes obtained using NF-HEDM. It can be seen that most of centroids 

Quantitatively, Fig.~\ref{fig:nf-hedm-cmp}(b) shows the distance between grains imaged using pseudo-Voigt FF-HEDM and NF-HEDM; the mean and median distances are 19.9~$\mu$m and 15.3~$\mu$m, respectively. 
Similarly, Fig.~\ref{fig:nf-hedm-cmp}(c) shows the distance between grains imaged using \proj{} and NF-HEDM.
The mean and median distances are 17.0~$\mu$m and 13.2~$\mu$m.
It can be seen that the results from \proj{} are, on average, 15\% better than pseudo-Voigt fitting.
The markers in Fig.~\ref{fig:nf-hedm-cmp}(b-c) are colored according to mean of the difference in position of expected and observed diffraction spots for each grain for pseudo-Voigt and \proj{}, respectively.
Here too, \proj{} performs better than pseudo-Voigt fitting by 28.6\%.
Internal Angle, another measure of quality of reconstructions, is mean of the angle between expected and observed diffraction-vectors for each grain.
The size of markers in Fig.~\ref{fig:nf-hedm-cmp}(b-c) is directly proportional to Internal Angle of the respective grain with \proj{} performing better than pseudo-Voigt fitting by 13\%.

\subsection{Computational Efficiency}\label{sec2:eva-comp}
Comparison of the reconstructed grain characteristics obtained with \proj{} with those obtained with conventional methods reveals no noticeable difference. 
However, \proj{} is much faster.
Our highly optimized implementation of 2D pseudo-Voigt fitting, coded in the C programming language, takes about 400 core-seconds to process a dataset of \num{800000} peaks on an 2.6 GHz, four-core, Intel Xeon server processor. 
On the same platform, \proj{} takes less than 7 core-seconds to process the dataset, a speedup of 57$\times$. 
As it is an out-of-the-box solution to run \proj{} on a GPU with any deep learning framework~(i.e., no extra efforts needed to program \proj{} for GPU), we also evaluate \proj{} on a NVIDIA V100 GPU. 

Analysis of the dataset takes only 280ms, for a speedup of more than 350$\times$ relative to the pseudo-Voigt fitting code on a quad-core server CPU~(to the best of our knowledge, there is no GPU-accelerated 2D pseudo-Voigt fitting implementation available so far).
If no server-class GPU is available near the experiment facility, \proj{} on a desktop with an affordable gaming NVIDIA RTX 2080Ti card only takes about 400ms, a speedup of 250$\times$ relative to running conventional pseudo-Voigt fitting on a high-end workstation CPU. 
We note that the dataset that we used for our evaluation is small, having been collected at only every 0.25$^\circ$ (1440 images for 360$^\circ$).
If we collect with step size of 0.01$^\circ$ (36000 images for 360$^\circ$) to assure better angular resolution in peak coordinates, the conventional method will take hundreds of hours to process all peaks while \proj{} can do it within an hour. 

\section{Ablation Study}\label{sec:ablation}
We describe experiments that we used to study the effectiveness of the data augmentation method described in \S\ref{sec2:data-aug} and the non-local self-attention block in our \proj{} architecture design. 

\subsection{Non-local Attention}\label{sec2:atten-ablation}
\begin{figure}
\centering
\includegraphics[width=.8\linewidth]{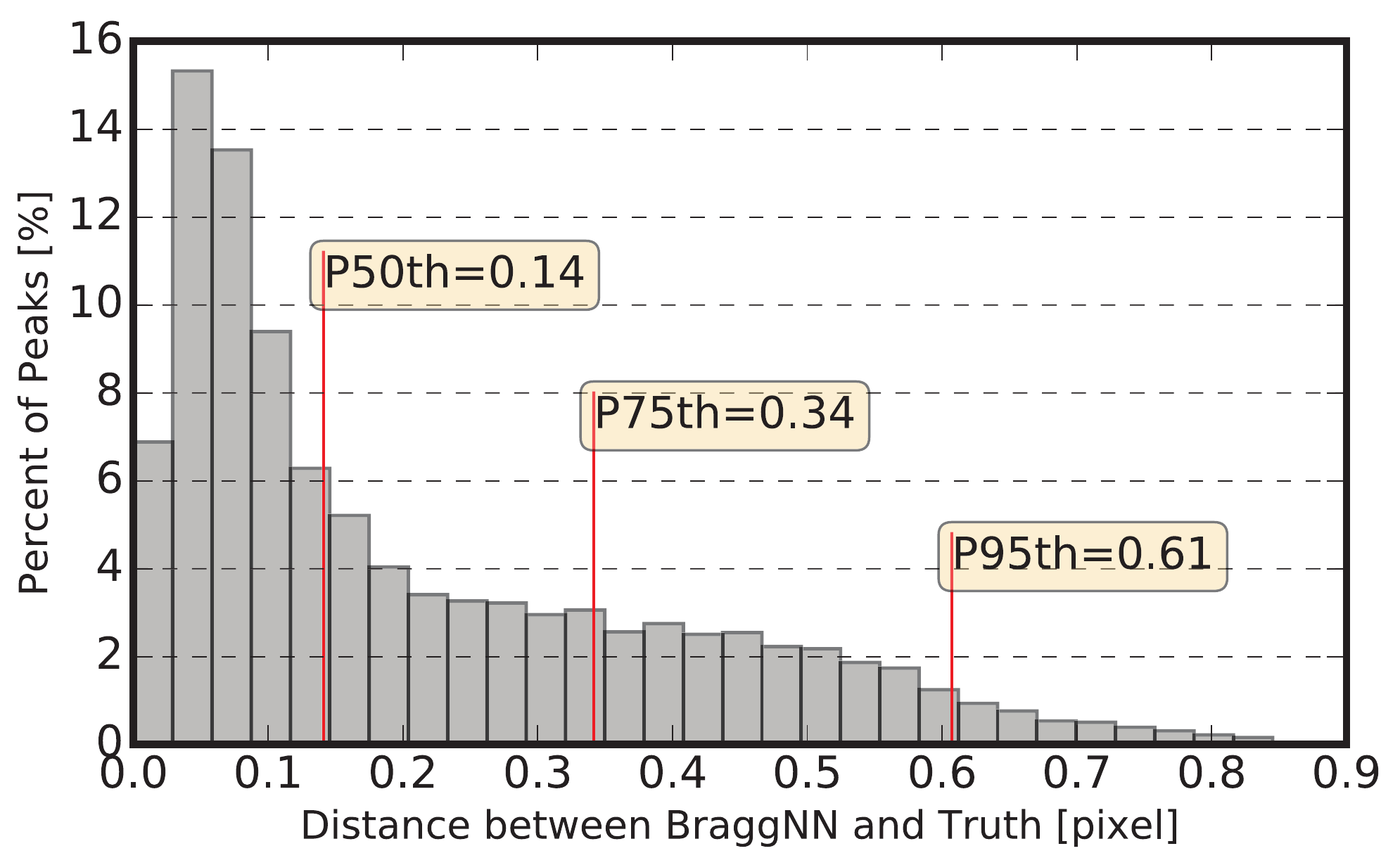}
\caption{Distribution of difference between peak positions located by \proj{} without the self-non-local self-attention block and conventional pseudo-Voigt fitting.}
\label{fig:wo-nlb}
\end{figure}

We used a non-local self-attention block on the feature maps of the first CNN layer for \proj{} in order to capture global information on the input patch of peak. 
The intuition behind this is that a global view at the early layer can help CNN layers better extract feature representation in the latent space for fully-connected layers to better approximate its center~\citep{wang2018non}. 
Here, we conduct an ablation study to show its effectiveness. 
We train two models, one with attention block one without, using the same datasets, i.e., attention block is the only difference, and then we evaluate their estimation accuracy. 
Fig.~\ref{fig:wo-nlb} shows the distribution of deviations. 
It is clear that both the 50th and 75th percentile deviations are more than 20\% worse than Fig.~\ref{fig:peak-diff}(c) where \proj{} has the non-local self-attention block, the 95th percentile is about 15\% worse. 

\subsection{Data Augmentation}
\label{appsec2:augm-ablation}
\begin{figure}
\centering
\includegraphics[width=\linewidth]{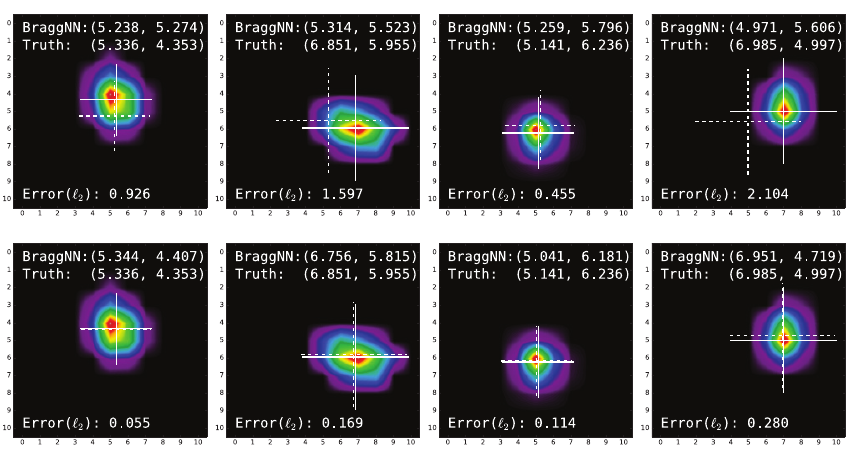}
% \vspace{-1cm}
\caption{Peaks located by \proj{} when peaks is deviated from geometric center. The error is the Euclidean distance~($\ell_2$) between truth~(solid line cross) and \proj{} prediction~(dotted line cross). Upper row: using \proj{} trained without data augmentation. Bottom row: using \proj{} trained with data augmentation for the same peaks as the upper row. }
\label{fig:devi-demo}
\end{figure}

We presented a novel data augmentation method to prevent model over-fitting and to address inaccurate patch cropping using the connect component in the model inference phase.
In order to study its effectiveness, we trained \proj{} on a simulation dataset with and without augmentation. 
When trained with augmentation, we use an interval of $\left[-1,1\right]$ for both $m$ and $n$. 
Fig.~\ref{fig:devi-demo} demonstrates three arbitrarily selected cases in our test dataset where the computed peak location deviated from the corresponding patch's geometric center~(i.e., (5, 5) for a $11\times11$ pixel patch) in different directions.
We can see from the demonstration that \proj{} is able to locate the peak values precisely even when the peak is deviated from the geometric center. 

In order to quantitatively evaluate the effectiveness of data augmentation, we sample $m$ and $n$ independently from \{-1, 0, 1\} when preparing our test dataset to mimic imperfect patch cropping. 
That is, only $1/3 \times 1/3=1/9$ of the patches have maxima at the geometric center. 

\begin{figure}
\centering
\includegraphics[width=\linewidth]{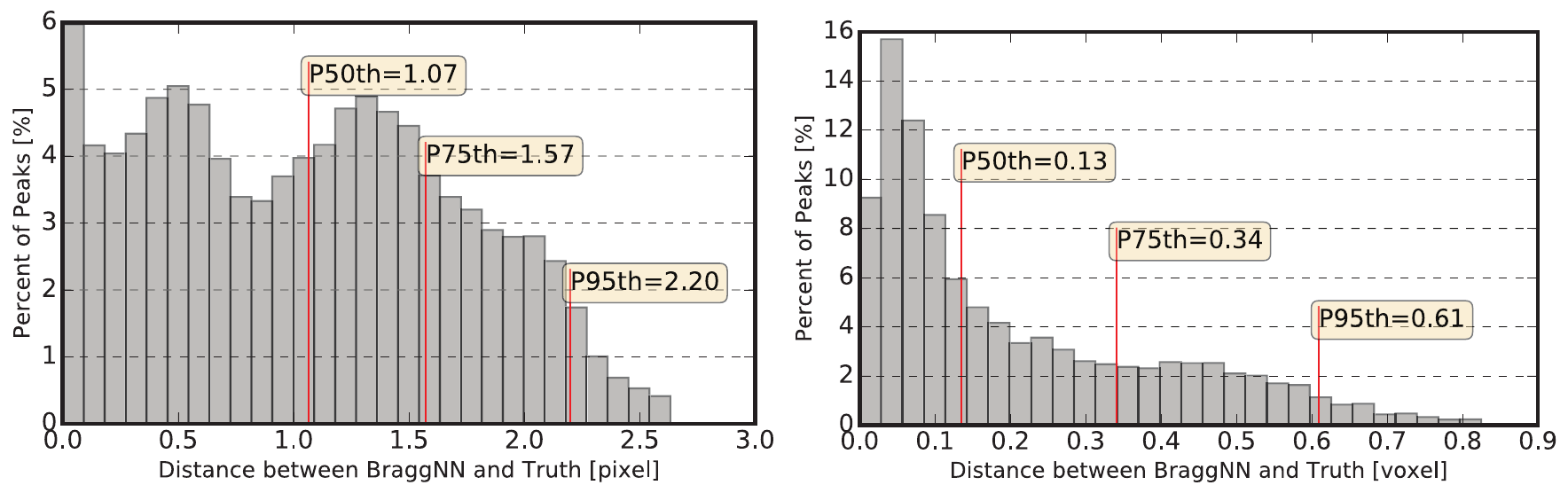}
% \vspace{-1cm}
\caption{Distribution of difference between peak position located by \proj{} (\texttt{right}: with~($m,n\in \{-1, 0, 1\}$) or \texttt{left}: without data augmentation) and conventional pseudo-Voigt fitting. $P_{nth}$ denotes the $n_{th}$ percentile.}
\label{fig:aug-ablation-exp}
\end{figure}

Fig.~\ref{fig:aug-ablation-exp} compares the prediction error on the test dataset in a statistical way. 
Comparing~Fig.~\ref{fig:aug-ablation-exp}(a) with Fig.~\ref{fig:aug-ablation-exp}(b), 
we see clear improvement when augmentation is applied for model training. 
The 50th, 75th, and 95th percentile errors are all reduced to about 20\% of those obtained when \proj{} is trained without data augmentation: a five times improvement.

\section{Conclusions and Future work}
We have described \proj{}, the first machine learning-based method for precisely characterizing Bragg diffraction peaks in HEDM images.
When compared with conventional 2D pseudo-Voigt fitting and higher resolution \nfhedm{}, \proj{}-localized peak-based reconstruction is within acceptable deviation while running more than 50$\times$ faster on a CPU and up to 1000$\times$ faster on a GPU. 
The speedup is important for high-resolution, high-throughput, and latency-sensitive applications, including real-time analysis and experiment steering~(e.g., searching area of interest for multi-scale image).

As for future work, we plan to extend \proj{} and apply a deep learning-based object localization technique directly to diffraction frames to (1) avoid labelling the connection component, (2) deal with dense peak diffraction where peaks are partially overlapped~---~a problem for which the conventional methods have exponential complexity, while the deep learning-based method has sub-linear complexity, and (3) generalize the model to work with 3D peaks for deformed grains. 

\section*{Acknowledgements}
This material was based upon work supported by the U.S. Department of Energy, Office of Science, under contract DE-AC02-06CH11357 and Office of Basic Energy Sciences under Award Number FWP-35896.
This research used resources of the Argonne Leadership Computing Facility, which is a DOE Office of Science User Facility supported under Contract DE-AC02-06CH11357. The authors would like to thank Paul Shade (Air Force Research Laboratory) for providing access to the experimental data set used as an example in this paper.

\bibliographystyle{unsrtnat}
\bibliography{references}  

\begin{thebibliography}{23}
\providecommand{\natexlab}[1]{#1}
\providecommand{\url}[1]{\texttt{#1}}
\expandafter\ifx\csname urlstyle\endcsname\relax
  \providecommand{\doi}[1]{doi: #1}\else
  \providecommand{\doi}{doi: \begingroup \urlstyle{rm}\Url}\fi

\bibitem[Park et~al.(2017)Park, Zhang, Kenesei, Wong, Li, and
  Almer]{park2017far}
Jun-Sang Park, Xuan Zhang, Peter Kenesei, Su~Leen Wong, Meimei Li, and Jonathan
  Almer.
\newblock Far-field high-energy diffraction microscopy: A non-destructive tool
  for characterizing the microstructure and micromechanical state of
  polycrystalline materials.
\newblock \emph{Microscopy Today}, 25\penalty0 (5):\penalty0 36--45, 2017.

\bibitem[Naragani et~al.(2017)Naragani, Sangid, Shade, Schuren, Sharma, Park,
  Kenesei, Bernier, Turner, and Parr]{NARAGANI201771}
Diwakar Naragani, Michael~D. Sangid, Paul~A. Shade, Jay~C. Schuren, Hemant
  Sharma, Jun-Sang Park, Peter Kenesei, Joel~V. Bernier, Todd~J. Turner, and
  Iain Parr.
\newblock Investigation of fatigue crack initiation from a non-metallic
  inclusion via high energy x-ray diffraction microscopy.
\newblock \emph{Acta Materialia}, 137:\penalty0 71 -- 84, 2017.

\bibitem[Bernier et~al.(2020)Bernier, Suter, Rollett, and Almer]{bernier2020}
Joel~V. Bernier, Robert~M. Suter, Anthony~D. Rollett, and Jonathan~D. Almer.
\newblock High-energy x-ray diffraction microscopy in materials science.
\newblock \emph{Annual Review of Materials Research}, 50\penalty0 (1):\penalty0
  395--436, 2020.

\bibitem[Wang et~al.(2020)Wang, Li, Li, Wang, Xu, Xue, Bai, and
  Wang]{WANG2020152534}
Shengjie Wang, Shilei Li, Runguang Li, Youkang Wang, Ning Xu, Fei Xue, Guanghai
  Bai, and Yan-Dong Wang.
\newblock Microscopic stress and crystallographic orientation of hydrides
  precipitated in zr-1nb-0.01cu cladding tube investigated by high-energy x-ray
  diffraction and ebsd.
\newblock \emph{Journal of Nuclear Materials}, 542:\penalty0 152534, 2020.

\bibitem[Sharma et~al.(2012{\natexlab{a}})Sharma, Huizenga, and
  Offerman]{Sharma2012ASetup}
Hemant Sharma, Richard~M. Huizenga, and S.~Erik Offerman.
\newblock {A fast methodology to determine the characteristics of thousands of
  grains using three-dimensional X-ray diffraction. I. Overlapping diffraction
  peaks and parameters of the experimental setup}.
\newblock \emph{Journal of Applied Crystallography}, 45\penalty0 (4):\penalty0
  693--704, 8 2012{\natexlab{a}}.
\newblock ISSN 00218898.
\newblock \doi{10.1107/S0021889812025563}.

\bibitem[Sharma et~al.(2012{\natexlab{b}})Sharma, Huizenga, and
  Offerman]{Sharma2012AGrains}
Hemant Sharma, Richard~M. Huizenga, and S.~Erik Offerman.
\newblock {A fast methodology to determine the characteristics of thousands of
  grains using three-dimensional X-ray diffraction. II. Volume, centre-of-mass
  position, crystallographic orientation and strain state of grains}.
\newblock \emph{Journal of Applied Crystallography}, 45\penalty0 (4):\penalty0
  705--718, 8 2012{\natexlab{b}}.
\newblock ISSN 00218898.
\newblock \doi{10.1107/S0021889812025599}.

\bibitem[Streiffer et~al.(2015)Streiffer, Vogt, Evans, et~al.]{APSU}
Stephen Streiffer, Stefan Vogt, Paul Evans, et~al.
\newblock Early science at the upgraded {Advanced Photon Source}.
\newblock Technical report, Argonne National Laboratory, October 2015.

\bibitem[Hornik(1991)]{hornik1991approximation}
Kurt Hornik.
\newblock Approximation capabilities of multilayer feedforward networks.
\newblock \emph{Neural networks}, 4\penalty0 (2):\penalty0 251--257, 1991.

\bibitem[Cheng et~al.(2017)Cheng, Wang, Zhou, and Zhang]{cheng2017survey}
Yu~Cheng, Duo Wang, Pan Zhou, and Tao Zhang.
\newblock A survey of model compression and acceleration for deep neural
  networks.
\newblock \emph{arXiv preprint arXiv:1710.09282}, 2017.

\bibitem[Fiorio and Gustedt(1996)]{fiorio1996two}
Christophe Fiorio and Jens Gustedt.
\newblock Two linear time union-find strategies for image processing.
\newblock \emph{Theoretical Computer Science}, 154\penalty0 (2):\penalty0
  165--181, 1996.

\bibitem[Wu et~al.(2005)Wu, Otoo, and Shoshani]{wu2005optimizing}
Kesheng Wu, Ekow Otoo, and Arie Shoshani.
\newblock Optimizing connected component labeling algorithms.
\newblock In \emph{Medical Imaging 2005: Image Processing}, volume 5747, pages
  1965--1976. International Society for Optics and Photonics, 2005.

\bibitem[Paszke et~al.(2019)Paszke, Gross, Massa, Lerer, Bradbury, Chanan,
  Killeen, Lin, Gimelshein, Antiga, et~al.]{paszke2019pytorch}
Adam Paszke, Sam Gross, Francisco Massa, Adam Lerer, James Bradbury, Gregory
  Chanan, Trevor Killeen, Zeming Lin, Natalia Gimelshein, Luca Antiga, et~al.
\newblock Pytorch: An imperative style, high-performance deep learning library.
\newblock In \emph{Advances in neural information processing systems}, pages
  8026--8037, 2019.

\bibitem[Kingma and Ba(2014)]{kingma2014adam}
Diederik~P Kingma and Jimmy Ba.
\newblock Adam: A method for stochastic optimization.
\newblock \emph{arXiv preprint arXiv:1412.6980}, 2014.

\bibitem[Cubuk et~al.(2018)Cubuk, Zoph, Man{\'{e}}, Vasudevan, and
  Le]{AutoAugment}
Ekin~Dogus Cubuk, Barret Zoph, Dandelion Man{\'{e}}, Vijay Vasudevan, and
  Quoc~V. Le.
\newblock Auto{A}ugment: Learning augmentation policies from data.
\newblock \emph{CoRR}, abs/1805.09501, 2018.
\newblock URL \url{http://arxiv.org/abs/1805.09501}.

\bibitem[Shorten and Khoshgoftaar(2019)]{shorten2019survey}
Connor Shorten and Taghi~M Khoshgoftaar.
\newblock A survey on image data augmentation for deep learning.
\newblock \emph{Journal of Big Data}, 6\penalty0 (1):\penalty0 60, 2019.

\bibitem[Krizhevsky(2009)]{krizhevsky2009learning}
Alex Krizhevsky.
\newblock Learning multiple layers of features from tiny images.
\newblock Technical report, University of Toronto, 2009.

\bibitem[Deng et~al.(2009)Deng, Dong, Socher, Li, Li, and
  Fei-Fei]{deng2009imagenet}
Jia Deng, Wei Dong, Richard Socher, Li-Jia Li, Kai Li, and Li~Fei-Fei.
\newblock Image{N}et: A large-scale hierarchical image database.
\newblock In \emph{IEEE Conference on Computer Vision and Pattern Recognition},
  pages 248--255. Ieee, 2009.

\bibitem[Zhang et~al.(2016)Zhang, Bengio, Hardt, Recht, and
  Vinyals]{zhang2016understanding}
Chiyuan Zhang, Samy Bengio, Moritz Hardt, Benjamin Recht, and Oriol Vinyals.
\newblock Understanding deep learning requires rethinking generalization.
\newblock \emph{arXiv preprint arXiv:1611.03530}, 2016.

\bibitem[Shade et~al.(2016)Shade, Menasche, Bernier, Kenesei, Park, Suter,
  Schuren, and Turner]{ShadeGold}
Paul~A. Shade, David~B. Menasche, Joel~V. Bernier, Peter Kenesei, Jun-Sang
  Park, Robert~M. Suter, Jay~C. Schuren, and Todd~J. Turner.
\newblock {Fiducial marker application method for position alignment of {\it in
  situ} multimodal X-ray experiments and reconstructions}.
\newblock \emph{Journal of Applied Crystallography}, 49\penalty0 (2):\penalty0
  700--704, Apr 2016.

\bibitem[Goodfellow et~al.(2016)Goodfellow, Bengio, and
  Courville]{goodfellow2016deep}
Ian Goodfellow, Yoshua Bengio, and Aaron Courville.
\newblock \emph{Deep Learning}.
\newblock MIT press, 2016.

\bibitem[Turner et~al.(2016)Turner, Shade, Bernier, Li, Schuren, Lind, Lienert,
  Kenesei, Suter, Blank, et~al.]{turner2016combined}
Todd~J Turner, Paul~A Shade, Joel~V Bernier, Shiu~Fai Li, Jay~C Schuren,
  Jonathan Lind, Ulrich Lienert, Peter Kenesei, Robert~M Suter, Basil Blank,
  et~al.
\newblock Combined near-and far-field high-energy diffraction microscopy
  dataset for ti-7al tensile specimen elastically loaded in situ.
\newblock \emph{Integrating Materials and Manufacturing Innovation}, 5\penalty0
  (1):\penalty0 94--102, 2016.

\bibitem[Sharma(2020)]{Sharma2020}
Hemant Sharma.
\newblock Midas software package.
\newblock \url{https://github.com/marinerhemant/MIDAS}, 2020.

\bibitem[Wang et~al.(2018)Wang, Girshick, Gupta, and He]{wang2018non}
Xiaolong Wang, Ross Girshick, Abhinav Gupta, and Kaiming He.
\newblock Non-local neural networks.
\newblock In \emph{IEEE Conference on Computer Vision and Pattern Recognition},
  pages 7794--7803, 2018.

\end{thebibliography}

\section*{Government License}
The submitted manuscript has been created by UChicago Argonne, LLC, Operator of Argonne National Laboratory (``Argonne''). Argonne, a U.S.\ Department of Energy Office of Science laboratory, is operated under Contract No.\ DE-AC02-06CH11357. The U.S.\ Government retains for itself, and others acting on its behalf, a paid-up nonexclusive, irrevocable worldwide license in said article to reproduce, prepare derivative works, distribute copies to the public, and perform publicly and display publicly, by or on behalf of the Government.  The Department of Energy will provide public access to these results of federally sponsored research in accordance with the DOE Public Access Plan. http://energy.gov/downloads/doe-public-access-plan.

\end{document}